\pdfoutput=1

\documentclass[11pt]{article}

\usepackage{EMNLP2023}

\usepackage{times}
\usepackage{latexsym}
\usepackage{multicol}
\usepackage{booktabs}
\usepackage[fleqn]{amsmath}
\usepackage{multirow}
\usepackage[normalem]{ulem}
\usepackage{hyperref}
\usepackage{graphicx}
\usepackage{subfigure}
\usepackage{fixltx2e}
\usepackage{algorithm}
\usepackage{algorithmic}
\usepackage{setspace}
\usepackage{soul}
\usepackage{amsmath}
\newcommand{\ie}{\textit{i.e.}}
\newcommand{\eg}{\textit{e.g.}}

\usepackage[T1]{fontenc}

\usepackage[utf8]{inputenc}

\usepackage{microtype}

\usepackage{inconsolata}

\title{History-Aware Conversational Dense Retrieval}

\author{Fengran Mo$^1$, Chen Qu$^{2}$, Kelong Mao$^{3}$, Tianyu Zhu$^{1,4}$\thanks{$^{*}$This work was done when Tianyu Zhu and Zhan Su were working at University of Montreal.}, Zhan Su$^{1,5*}$, Kaiyu Huang$^{6\dagger}$, Jian-Yun Nie$^{1}$\thanks{$^{\dagger}$Corresponding authors.}\\
$^1$University of Montreal, Quebec, Canada; 
$^2$University of Massachusetts Amherst, USA \\ 
$^3$Renmin University of China; 
$^4$Beihang University, China \\
$^5$University of Copenhagen, Denmark; 
$^6$Beijing Jiaotong University, China \\
\texttt{fengran.mo@umontreal.ca, kyhuang@bjtu.edu.cn, nie@iro.umontreal.ca} \\
}

\begin{document}
\maketitle
\begin{abstract}
Conversational search facilitates complex information retrieval by enabling multi-turn interactions between users and the system. Supporting such interactions requires a comprehensive understanding of the conversational inputs to formulate a good search query based on historical information. In particular, the search query should include the relevant information from the previous conversation turns.
However, current approaches for conversational dense retrieval primarily rely on fine-tuning a pre-trained ad-hoc retriever using the whole conversational search session, which can be lengthy and noisy. 
Moreover, existing approaches are limited by the amount of manual supervision signals in the existing datasets.
To address the aforementioned issues, we propose a \textbf{H}istory-\textbf{A}ware \textbf{Conv}ersational \textbf{D}ense \textbf{R}etrieval (HAConvDR) system, which incorporates two ideas: context-denoised query reformulation and automatic mining of supervision signals based on the actual impact of historical turns.
Experiments on two public conversational search datasets demonstrate the improved history modeling capability of HAConvDR, in particular for long conversations with topic shifts.
\end{abstract}

\section{Introduction}
Conversational search is expected to be the next generation of search engines~\cite{gao2022neural}. It aims to satisfy complex user information needs via multi-turn interactions between a user and the system.
In single-turn ad-hoc search, users typically employ stand-alone queries to convey their information requirements~\cite{bajaj2016ms} in a brief and clearly-expressed manner.
In conversational search, however, queries are usually context-dependent, which highlights the necessity of understanding the search intent within the conversational context.

To uncover the user's information need, conversational query rewriting (CQR)~\citep{yu2020few,wu2022conqrr,mo2023convgqr} employs human-rewritten queries to train a rewriting model that generates de-contextualized queries.
However, obtaining large-scale manual annotations for this purpose is challenging in practice. Besides, CQR models cannot be directly optimized for the downstream retrieval task~\cite{wu2022conqrr,mo2023convgqr}. 

In comparison, a more desirable approach is to perform end-to-end conversational dense retrieval (CDR) by training a query encoder that incorporates conversation history~\cite{qu2020open,yu2021few}. Since human annotations are usually not available to indicate which previous conversation turns are relevant to the current query, a common practice is to utilize all historical turns to reformulate the current query as the input to the model.

\begin{figure}[t]
\centering
\includegraphics[width=1\linewidth]{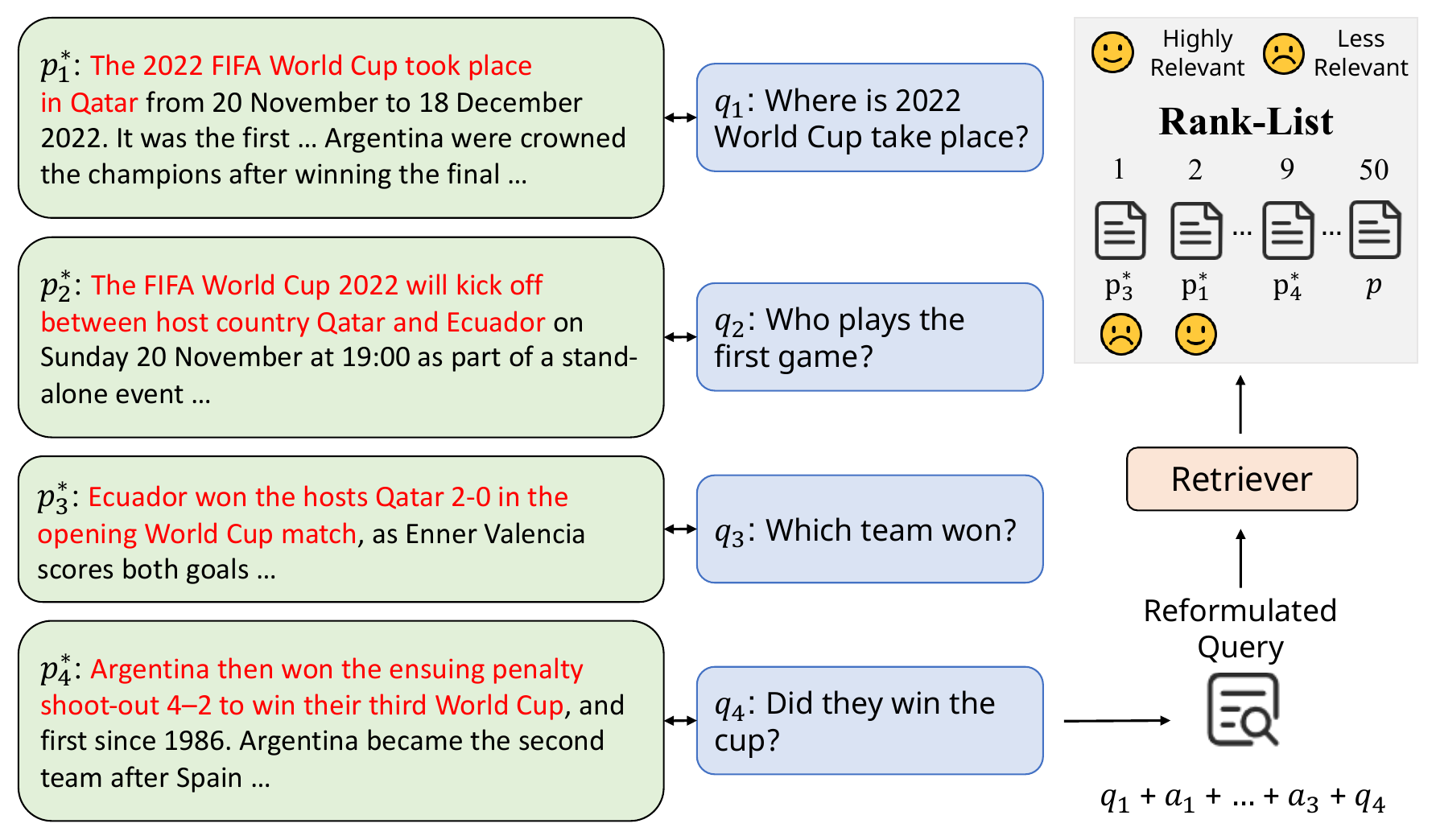}
    \caption{Illustration of shortcut history dependency -- passages addressing historical information needs $p_3^*$ can be ranked higher than those addressing the current information need $p_4^*$, due to the noise in the reformulated query. The highly relevant passage $p_1^*$ could be served as PRF. The red text denotes the gold answer $a_i$ in $p_i^*$.}
	\label{fig: example}
\vspace{-2ex}
\end{figure}

However, the conversation history can be lengthy and often includes a substantial amount of noise, \ie, historical turns that are irrelevant to the current query. 
Despite the observation~\cite{adlakha2022topiocqa} that conversational sessions often center around a specific topic (\eg, sports), it is worth noting that different turns may focus on different aspects (\eg, match results, or player statistics). Some of them are relevant to the current turn, while others may not. This is especially the case when conversations are long.
This problem can give rise to the issue of shortcut history dependency~\cite{kim2022saving,fang2022open}, i.e. the reformulated query depends excessively on the historical turns while neglecting the current query. 
We illustrate this issue by an example in Figure~\ref{fig: example}. Given the current query $q_4$, instead of retrieving the passage $p_4^*$ (addressing the current information need) in top-ranked positions, the retriever ranks $p_3^*$ (addressing historical information needs) higher than $p_4^*$.

To tackle the aforementioned challenge, we put forward \textbf{HAConvDR}, a new \textbf{H}istory-\textbf{A}ware \textbf{Conv}ersational \textbf{D}ense \textbf{R}etrieval method, aiming to leverage the useful information from the history as much as possible to reformulate the current query. Our approach consists of two prongs of enhancements as detailed in the following sections. 

The first prong is to incorporate an explicit denoising mechanism into the model training process so that the model is less affected by the noisy history while being history-aware.
To achieve a similar purpose, recent studies~\cite{mao2022curriculum,mao2023learning,mo2023learning} typically assess whether a historical turn is relevant to the current turn based on the historical query. However, these approaches are inherently lacking because historical queries alone are often not sufficient to fully cover the historical context. To address this shortcoming, we additionally leverage the passages associated with historical queries to better evaluate the intent of a historical turn. Specifically, we use a pseudo-labeling approach to assess the relevance and usefulness of the historical turns -- whether they contribute to improving the retrieval effectiveness of the current query. We then retain the relevant historical turns for context-denoised query reformulation.

The second prong is to mine additional supervision signals to further alleviate the pitfall of shortcut history dependency. Despite having context-denoised queries, a single ground-truth passage (given by the dataset) is often indirect and insufficient to guide the training of conversational retrieval due to the remaining noise in the formulated query.
Thus, mining additional supervisions, either positive~\cite{mao2022convtrans} or negative~\cite{kim2022saving}, can enhance the original supervision signal and reduce the negative impact by the distractors in the conversation history.
Different from the aforementioned work that acquires additional supervisions by human annotation or retrieval, we mine pseudo positive and hard negative supervisions from the conversation history based on the same relevance judgment of historical turns used for query reformulation. Intuitively, among the top-ranked historical ground-truth passages in Figure~\ref{fig: example}, some of them can be highly relevant to the current query, which resembles the pseudo relevant documents in Pseudo Relevance Feedback (PRF)~\cite{Xu1996QueryEU}, while others are less relevant and can serve as hard negatives for training. These additional supervisions enable the model to be aware of the usefulness or harmfulness of historical ground-truth passages and leverage them in a history-aware contrastive learning process. 

We carry out extensive experiments on two conversational search datasets to test the effectiveness of HAConvDR. The results show that our method outperforms most existing strong baselines, demonstrating how relevance judgments of historical turns can benefit conversational retrieval.

Our contributions are summarized as follows: (1) We propose HAConvDR to train a history-aware conversational dense retriever by using the ground-truth passage from historical turns as additional supervision signals.
(2) We conduct pseudo relevance judgment on selecting historical turns to denoise the context for query reformulation, whose results are the foundation of mining additional supervision signals.
(3) We demonstrate the effectiveness of HAConvDR by outperforming different types of strong baselines on two public datasets.
A series of analyses are conducted to understand how historical ground-truth passages work well to solve the conversation with lots of topic shifts.

\section{Related Work}
\textbf{Conversational Query Reformulation.}
This approach aims to reformulate an explicit query via training a CQR model. Typical methods include query rewriting~\cite{yu2020few,lin2020conversational,vakulenko2021question,qian2022explicit,mao2023large,mao2023search} and query expansion~\cite{2020Making,voskarides2020query}, which aim to mimic human query rewriting or selection of useful terms from historical context for expansion. However, the manual annotations needed for training are difficult to obtain in practice and the human-rewritten queries might not necessarily be the optimal search queries~\cite{wu2022conqrr,mo2023convgqr}. Some recent studies~\cite{ye2023enhancing,mao2023large,jang2023itercqr} leverage large language models (LLMs) to generate reformulated queries via prompting but the generated queries are not optimized for search.\\

\noindent \textbf{Conversational Dense Retrieval.}
Another research direction is to perform conversational dense retrieval, which leverages conversational search data to fine-tune a well-trained ad-hoc retriever. Existing studies~\cite{yu2021few,lin2021contextualized,mao2022convtrans,mo2024convsdg,chen2024generalizing} usually focus on few-shot scenarios or rely on external resources, but without context denoising. On context denoising, some recent work~\citep{mao2022curriculum,mao2023learning,mo2023learning,mao2024chatretriever} designs sophisticated mechanisms to enhance the denoising ability explicitly and implicitly for the models. However, they do not take into account historical feedback.
To perform context-denoising more effectively, our method explicitly selects the useful historical turns, as well as their ground-truth passage via pseudo relevant judgment before model training. \\

\noindent \textbf{Supervision Signals in Dense Retrieval.}
\citet{robinson2021contrastive} demonstrates that sufficient supervision signals, either positive or negative (especially hard negatives), are important for contrastive learning. 
For dense retrieval, hard negatives are usually mined by BM25~\cite{karpukhin2020dense} or a vanilla backbone model~\cite{xiong2020approximate}. In the conversational scenario,~\citet{kim2022saving} uses the CQR model to construct hard negatives and~\citet{mao2022curriculum} relies on human annotators to generate augmented positives, but the amount of generated data is limited. Differently, our method leverages additional supervision signals from the historical ground-truth passages to enhance the model's history-awareness (\eg, enjoying the efficiency and avoiding the harmfulness).

\begin{figure*}[t]
\centering
\includegraphics[width=1\linewidth]{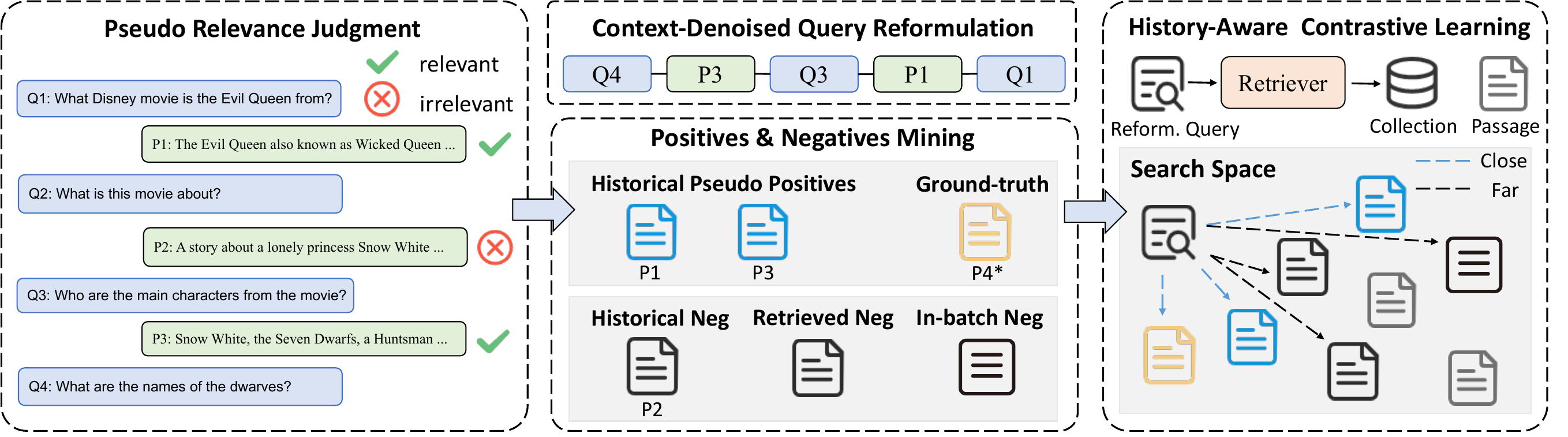}
	\caption{Overview of HAConvDR. The first stage (left) is to conduct pseudo relevance judgment (PRJ) between the current query and each historical turn. Based on the PRJ results, the second stage (middle) is to perform context-denoised query reformulation and positive and negative supervision signals mining. The third stage (right) is to conduct conversational dense retrieval training with history-aware contrastive learning.}
\label{fig: overview}
\vspace{-2ex}
\end{figure*}

\section{Methodology}
\label{sec: method}

\subsection{Task Definition}
\label{subsec: task}

We are given a conversation session that contains the current query $q_n$, and $n-1$ historical turns preceding $q_n$. The $i$-th historical turn is denoted as $(q_i, p_i^*)$, where $q_i$ is a historical query and $p_i^*$ is the historical ground-truth passage corresponding to $q_i$. Our task is to retrieve the passage $p_n^*$ from a passage collection $\mathcal{D}$ to satisfy the information need in $q_n$. Our utilization of historical ground-truth passages $\mathcal{P}_h = \{p_i^*\}_{i=1}^{n-1}$ is consistent with the settings adopted in previous work on conversational search~\cite{choi2018quac,Qu2019AttentiveHS}, i.e. we assume that the relevant passages for the previous turns are known. In some real-world applications, if $\mathcal{P}_h$ is not available, it can be replaced with a set of top-ranked passages for those turns. We discuss and analyze such adaptation in Sec.~\ref{subsec: Applicability of PRF} for generalizability.

\subsection{Method Overview}
As illustrated in Figure~\ref{fig: overview}, HAConvDR consists of three stages. The first stage is to generate the pseudo relevance judgments (PRJs) for historical turns by evaluating whether a given turn $(q_i, p_i^*)$ is relevant to the current query $q_n$. This is achieved by a pseudo-labeling approach presented in Sec.~\ref{subsec: PRJ}. In the second stage, we leverage the generated PRJs of historical turns for two purposes. The first purpose is to use the relevant historical turns to perform a context-denoised query reformulation (Sec.~\ref{subsec: query reformulation}), while the second purpose is to create additional positive and negative training pairs by leveraging historical passages according to their PRJs. Given the reformulated queries and the augmented training pairs from conversation history, we train a dense retriever based on dual-encoder by history-aware contrastive learning 
in the third stage (Sec.~\ref{subsec: cl}). We highlight that, in our approach, the conversation history is considered as a source of not only context information, but also 
supervision signals. 
We describe each stage as follows.

\subsection{Relevance Judgement for Historical Turns}
\label{subsec: PRJ}
A common practice to obtain a conversational dense retriever is to adapt models for ad-hoc retrieval to a conversational setting by concatenating the entire conversation history to the current query.
In theory, the attention mechanism within the backbone transformer should allow the adapted retriever to implicitly conduct history modeling.
In practice, however, the attention can be easily distracted by the irrelevant information in the conversation history.
Therefore, we argue that it is essential to judge whether a historical turn is relevant to the current turn as part of the history modeling process.

In the literature of information retrieval, \textit{relevance} is used to denote how well a \textit{document} meets the information need of a query. Here, we take the liberty of using the same term to describe whether a \textit{historical turn} is relevant to the current query.

Learning to judge the relevance of historical turns is non-trivial
because conversation datasets rarely contain such labels.
\citet{mo2023learning} addresses this issue by adopting a simple and effective 
approach based on real impact on retrieval to derive pseudo labels -- a historical query $q_i$ is judged relevant if concatenating it to the current query $q_n$ leads to an improved retrieval performance for $q_n$ (similar to selecting query expansion terms as in~\citet{cao2008selectinggood}). This pseudo-labeling approach is referred to as pseudo relevance judgment for historical turns.
Despite the direct association with the retrieval task, this approach is limited by the fact that it only considers the queries in the historical turns, while ignoring the relevant or retrieved passages for them. 
To leverage the full conversational IR context,  
we also include the corresponding passages for each historical turn in our approach. We use a similar idea to \citet{mo2023learning} to label if a relevant passage $p_i^*$  to a previous turn $i$ is also relevant to the current turn, by assessing the impact of it on retrieval when it is concatenated, together with the historical query, to the current query.

The algorithm is illustrated in Algorithm~\ref{alg: PRJ}. 
It divides the historical ground-truth passages $\mathcal{P}_h = \{p_i^*\}_{i=1}^{n-1}$ into two disjoint groups:
\begin{equation}
    \mathcal{P}^{+}_{h} = \{p^{*}_{j}\}_{j=1}, \quad \mathcal{P}^{-}_{h} = \{p^{*}_{k}\}_{k=1}
    \label{eq: prepos}
\end{equation}
where $\mathcal{P}^{+}_{h}$ denotes the \textit{relevant} passage group and  $\mathcal{P}^{-}_{h}$ denotes the \textit{irrelevant} passage group. 
For the use case where historical ground-truth passages are not available, we demonstrate that top-retrieved passages can serve as a substitute in Sec.~\ref{subsec: Applicability of PRF}. 

\begin{algorithm}[t]
\footnotesize
\setstretch{1.1}
\caption{Generating pseudo relevance judgments for historical turns}
\begin{algorithmic}[1]
\REQUIRE current query $q_n$, historical turn ($q_i$, $p_i^*$), retriever $\phi$, retrieval evaluation metric $\mathcal{M}$ \\
\STATE RankList-raw $\leftarrow \phi(q_n)$ \
\STATE RankList-reform. $\leftarrow \phi(q_n \circ q_i \circ p_i^*)$ \
\STATE Score-raw $\leftarrow \mathcal{M}(\text{RankList-raw})$ \
\STATE Score-reform. $\leftarrow \mathcal{M}(\text{RankList-reform.})$ \
\IF {Score-reform. > Score-raw} 
    \STATE $\text{PRJ}(q_n, (q_i, p_i^*)) \leftarrow \text{relevant}$ 
\ELSE
    \STATE $\text{PRJ}(q_n, (q_i, p_i^*)) \leftarrow \text{irrelevant}$ \\
\ENDIF
\STATE \textbf{Output} $\text{PRJ}(q_n, (q_i, p_i^*))$ \
\end{algorithmic}
\label{alg: PRJ}
\end{algorithm}

\subsection{Context-Denoised Query Reformulation}
\label{subsec: query reformulation}
Based on the PRJs of historical turns derived in Sec.~\ref{subsec: PRJ}, we reformulate the current query $q_n$ to obtain the context-denoised query $q^{r}_{n}$:
\begin{equation}
    \quad \quad \quad q^{r}_{n} = q_n \circ \cdots p_{i}^{*} \circ q_{i} \cdots
    \label{eq: reformulated query}
\end{equation}
where $q_i$ and $p_i^*$ are from relevant historical turns, and $\circ$ denotes concatenation. 

Since the reformulated query contains historical passages $\mathcal{P}_h^+$, a potential concern arises regarding the length of the reformulated query -- it might exceed the input length limitations of some pre-trained language models.
However, the analysis of the generated PRJ statistics, as presented later in Sec.~\ref{subsec: PRJ analysis}, reveals that only a small portion of historical turns are deemed relevant and used for query reformulation.
This indicates the practical feasibility of our approach.
Nonetheless, in our future work, we will consider developing a more sophisticated mechanism to make a more strict selection of relevant 
passages in $\mathcal{P}_h^+$.

\subsection{History-Aware Contrastive Learning}
\label{subsec: cl}

Contrastive learning is a prevalent approach to train dense retrievers~\cite{karpukhin2020dense}. This approach first projects queries and passages into an embedding space with dual encoders $\mathcal{F}_{Q}$ and $\mathcal{F}_{P}$. It then evaluates the relevance of any given pair of query and passage $(q, p)$ by taking the dot product similarity $\mathcal{S}(q, p) = \mathcal{F}_{Q}(q)^{T} \cdot \mathcal{F}_{P}(p)$. Finally, supervision signals are derived from the positive and negative passages so that the distance between a query and a relevant passage (positive pair) should be closer than that between the same query and an irrelevant passage (negative pair). These supervision signals are back-propagated to train the encoders.

In a research setting, for the current query $q_n$, the relevant passage (positive passage) is the ground-truth passage $p_n^*$ given by the dataset. For the irrelevant passages (negative passages), one option is to simply take the passages other than $p_n^*$ found in the same training batch. These negative passages are referred to as in-batch negatives, here denoted as $\mathcal{P}^{-}_{b}$. In addition to in-batch negatives, another commonly adopted approach is to leverage retrieved hard negatives $\mathcal{P}^{-}_{r}$~\cite{lin2021contextualized,kim2022saving,karpukhin2020dense}. One way to obtain such negatives is to use the top-ranked passages retrieved with $q_n$ by an off-the-shelf retriever (\eg, BM25) after removing $p_n^*$ (if present). Supervision signals generated from these retrieved negatives are believed to be more meaningful than those from in-batch negatives. The power of retrieved negatives suggests that the effectiveness of supervision signals could be heavily impacted by the quality and quantity of the positive and negative pairs. 

Given the insight that augmenting positive and negative pairs can boost retrieval performance, we propose to mine additional pairs to further enhance the contrastive learning process. For this very purpose, we found the PRJs of historical turns derived in Sec.~\ref{subsec: PRJ} come in handy. 

Intuitively, $\mathcal{P}_h^+$ contains historical passages from the historical turns that are deemed relevant to $q_n$. Although $\mathcal{P}_h^+$ may not directly address the information need of $q_n$, $\mathcal{P}_h^+$ helps enhance or complement $q_n$. We believe this relationship can serve as a proxy to claim a certain level of relevance between $\mathcal{P}_h^+$ and $q_n$. Therefore, we use $\mathcal{P}_h^+$ as \textit{pseudo positives}.
Similarly, passages in $\mathcal{P}_h^-$ are less relevant to $q_n$ as demonstrated by the irrelevant PRJs. So we use $\mathcal{P}_h^-$ as additional negatives. More importantly, $\mathcal{P}_h^-$ resembles retrieved negatives $\mathcal{P}^{-}_{r}$ in the sense that both are hard negatives that can generate more meaningful supervisions. We refer to $\mathcal{P}_h^-$ as \textit{historical hard negatives}. 

By leveraging these pseudo positives and historical hard negatives mined from the conversation history, we upgrade traditional contrastive learning to history-aware contrastive learning.
Formally, we denote the final positive and negative passages used for training as follows:
\begin{equation}
\begin{aligned}
    \mathcal{P}^{+}_{n} = \{p^{*}_{n}\} \cup \mathcal{P}^{+}_{h}, \quad &|\mathcal{P}^{+}_{n}| = N \\
    \mathcal{P}^{-}_{n} = \mathcal{P}^{-}_{b} \cup \mathcal{P}^{-}_{r} \cup \mathcal{P}^{-}_{h}, \quad &|\mathcal{P}^{-}_{n}| = M
\end{aligned}
\label{eq: pos neg sample}
\end{equation}
The final training objective is illustrated in Eq.~\ref{eq: loss}, where $p^{+}_{i} \in \mathcal{P}^{+}_{n}$ and $p^{-}_{j} \in \mathcal{P}^{-}_{n}$.
\begin{equation}
\mathcal{L} = \frac{1}{N}\sum^{N}_{i=1} \frac{e^{\mathcal{S}\left(q^{r}_{n}, p^{+}_{i}\right)}} {e^{\mathcal{S}\left(q^{r}_{n}, p^{+}_{i}\right)} + \sum^{M}_{j=1} e^{\mathcal{S}\left(q^{r}_{n}, p^{-}_{j}\right)}}
    \label{eq: loss}
\end{equation}

\begin{table*}[t]
    \centering
    \scalebox{0.9}{
    \begin{tabular}{clcccccccc}
    \toprule
        \multirow{2}{*}{Category} & \multirow{2}{*}{Method} & \multicolumn{4}{c}{TopiOCQA} & 
        \multicolumn{4}{c}{QReCC}\\ 
        \cmidrule{3-10}
        ~ & ~ & MRR & NDCG@3 & R@10 & R@100 & MRR & NDCG@3 & R@10 & R@100 \\ 
        \midrule  
        \multirow{8}{*}{CQR} & GPT2+WS & 12.6 & 12.0 & 22.0 & 33.1 & 33.9 & 30.9 & 53.1 & 72.9 \\
        ~ & QuReTeC & 11.2 & 10.5 & 20.2 & 34.3 & 35.0 & 32.6 & 55.0 & 72.9 \\
        ~ & CQE-sparse & 14.3 & 13.6 & 24.8 & 36.7 & 32.0 & 30.1 & 51.3 & 70.9 \\
        ~ & T5QR & 23.4 & 22.5 & 39.8 & 56.2 & 34.5 & 31.8 & 53.1 & 72.8 \\
        ~ & CONQRR & - & - & - & - & 41.8 & - & 65.1 & 84.7 \\
        ~ & ConvGQR & 25.6 & 24.3 & 41.8 & 58.8 & 42.0 & 39.1 & 63.5 & 81.8\\
        ~ & IterCQR & 26.3 & 25.1 & 42.6 & 62.0 & 42.9 & 40.2 & 65.5 & 84.1 \\
        ~ & LLM-Aided IQR & - & - & - & - & 43.9 & 41.3 & 65.6 & 79.6\\
        \midrule
        \multirow{5}{*}{CDR} & Conv-ANCE & 22.9 & 20.5 & 43.0 & 71.0 & 47.1 & \textbf{45.6} & 71.5 & 87.2\\
        ~ & InstructoR & 25.3 & 23.7 & 45.1 & 69.0 & 43.5 & 40.5 & 66.7 & 85.6 \\
        ~ & SDRConv & 26.1 & 25.4 & 44.4 & 63.2 & 47.3 & 43.6 & 69.8 & 88.4 \\
        ~ & ConvDR & 27.2 & 26.4 & 43.5 & 61.1 & 38.5 & 35.7 & 58.2 & 77.8 \\
        \cmidrule{2-10}
        ~ & HAConvDR (Ours) & \textbf{30.1}$^\dagger$ & \textbf{28.5}$^\dagger$ & \textbf{50.8}$^\dagger$ & \textbf{72.8}$^\dagger$ & \textbf{48.5}$^\dagger$ & \textbf{45.6} & \textbf{72.4}$^\dagger$ & \textbf{88.9}$^\dagger$ \\
        \bottomrule
     \end{tabular}}
     \caption{Performance of different dense retrieval methods on two datasets.
    $\dagger$ denotes significant improvements with t-test at $p<0.05$ over the main competitors, all CDR methods. \textbf{Bold} indicate the best results.}
     \label{table: Main Results}
\vspace{-2ex}
\end{table*}

\section{Experiments}
\label{sec: exp}
\noindent \textbf{Datasets} \quad We evaluate our methods on two widely-used conversation datasets. The first is the TopiOCQA~\cite{adlakha2022topiocqa} dataset that contains complex topic-switch phenomena within each conversational session. These sessions have the potential to conceal a wealth of supervision signals in historical turns.
The other dataset we use is QReCC~\cite{anantha2021open}, where most queries in a conversational session are on the same topic. 
The selection of the datasets assures we verify the model performance on conversations with different intrinsic characteristics and enables more informative analyses. 
The statistics and more details of the datasets are provided in Appendix~\ref{appendix: datasets}.\\ 

\noindent \textbf{Evaluation metrics} \quad For an adequate comparison with previous studies, we use four standard evaluation metrics: MRR, NDCG@3, Recall@10, and Recall@100 to evaluate the retrieval results. \\

\noindent \textbf{Baselines} \quad We compare our method with two lines of conversational search approaches. The first line (CQR) performs conversational query reformulation based on generative rewriter models and off-the-shelf retrievers, including PLM-based \texttt{GPT2+WS}~\cite{yu2020few}, \texttt{QuReTeC}~\cite{voskarides2020query}, \texttt{CQE-Sparse}~\cite{lin2021contextualized}, \texttt{T5QR}~\cite{lin2020conversational}, \texttt{CONQRR}~\cite{wu2022conqrr}, and \texttt{ConvGQR}~\cite{mo2023convgqr}, and LLM-based \texttt{IterCQR}~\cite{jang2023itercqr}, and \texttt{LLM-Aided IQR}~\cite{ye2023enhancing}.
The second line (CDR) conducts conversational dense retrieval based on ad-hoc search dense retrievers to learn the latent representation of the reformulated query, including \texttt{Conv-ANCE}~\cite{mao2023learning} using the original contrastive ranking loss, \texttt{InstructorR}~\cite{jin2023instructor} utilizing LLMs to predict the relevance score between the session and passages then conduct the training of the retriever, \texttt{ConvDR}~\cite{yu2021few}  relying also on human-rewritten queries as supervision signals and \texttt{SDRConv}~\cite{kim2022saving} that includes mining additional hard negatives. The LLM-based methods employ ChatGPT or LLaMa as backbone models. \\

\noindent \textbf{Implementation details} \quad 
The backbone model for conversational dense retriever training is ANCE~\cite{xiong2020approximate} and the dense retrieval is performed using Faiss~\citep{johnson2019billion}.
During training, we only update the parameters of the query encoder while keeping the passage encoder frozen.
The number of mined positives and negatives from historical turns can vary across different query turns. Instead of trying to utilize all of them, we randomly select one historical pseudo positive and one historical hard negative (along with the top retrieved hard negative) for each training instance to strike a balance between effectiveness and efficiency.
More details are provided in Appendix~\ref{appendix: Implementation} and our 
code.\footnote{\url{https://github.com/fengranMark/HAConvDR}}

\subsection{Main Results}
\label{subsec: Main Results}

The main evaluation results on TopiOCQA and QReCC datasets are reported in Table~\ref{table: Main Results}. 

We find that our method achieves a significantly better performance on both datasets compared with other methods on most metrics. In particular, it improves MRR by 10.7\% and NDCG@3 by 8.0\% on TopiOCQA over the second-best results ConvDR.
The superior effectiveness can be attributed to the following two aspects. (1) The context-denoised query reformulation and history-aware contrastive learning with mined supervision signals enhance the ranking ability of our HAConvDR. (2) Conversational dense retrieval tends to be more effective compared with conversational query rewriting pipelines, including those leveraging the powerful generation capacity of LLMs.
Besides, the improvements achieved over Conv-ANCE serve as additional validation of the effectiveness of exploiting supplementary supervision signals derived from ground-truth information of past interactions and confirm our underlying assumption.

Moreover, we find that performance improvements are more pronounced on TopiOCQA. This can be attributed to the characteristics of the datasets: the session context is longer in TopiOCQA, and contains more noise. This comparison indicates that our method has a greater potential for longer sessions with topic shifts. 
In contrast, the turns in QReCC are usually on the same topic, and the ground-truth passages of historical turns can also properly address the information needs of the current query. In such a situation, most previous turns can be relevant, making it less critical to select the relevant turns. 
Notice that TopiOCQA provides a better simulation of real-world scenarios, where a conversation (or search) session is expected to be on related but different topics. 
Our results demonstrate that our approach is better at addressing this practical situation. More analysis on this is provided in Sec.~\ref{subsec: PRJ analysis} and Sec.~\ref{subsec: impact of historical supervision signals}.

\subsection{Ablation Study}
\label{subsec: ablation}

\begin{table}[!t]
\centering
\small
\setlength{\tabcolsep}{4pt}{
\begin{tabular}{lcccc}
\toprule
& \multicolumn{2}{c}{TopiOCQA} & \multicolumn{2}{c}{QReCC} \\
\cmidrule(lr){2-3}\cmidrule(lr){4-5}
~ & {MRR} & {NDCG@3} & {MRR} & {NDCG@3} \\
\midrule
Ours & \textbf{30.1} & \textbf{28.5} & \textbf{48.5} & \textbf{45.6}\\
\quad \texttt{- hard neg.} & 28.2 & 26.6 & 47.8 & 44.7\\
\quad \texttt{- pse. pos.} & 26.8 & 25.3 & 46.8 & 44.1\\
\quad \texttt{- QR w/ PRJ} & 25.0 & 23.0 & 44.5 & 41.4\\
\bottomrule
\end{tabular}
}
\caption{Ablation study of different strategies.}
\label{table: ablation}
\vspace{-2ex}
\end{table}
Compared to the contrastive learning technique in conversational dense retrieval, our proposed method introduces two extra components, \ie, context-denoised query reformulation and history-aware contrastive signals comprising historical pseudo positives and historical hard negatives. To assess the effectiveness of these individual components, we conduct an ablation study and present the analysis in Table~\ref{table: ablation}.

We observe that, on both datasets, removing pseudo positives can cause a more pronounced performance degradation compared with removing hard negatives. This suggests that, although both hard negatives and pseudo positives are useful, the latter serves as a more effective supervision. This insight complements the currently prevalent studies on negative mining. On the other hand, we observe the decrease is more prominent on TopiOCQA, which is true for both removing hard negatives and pseudo positives. This can be attributed to the prevalence of topic-switch phenomena within the sessions in TopiOCQA, where historical supervision can and should be leveraged to boost performance as illustrated in our approach.

\subsection{Investigation of PRJs of Historical Turns}
\label{subsec: PRJ analysis}
\begin{figure}[!t]
\centering
\includegraphics[width=0.85\linewidth]{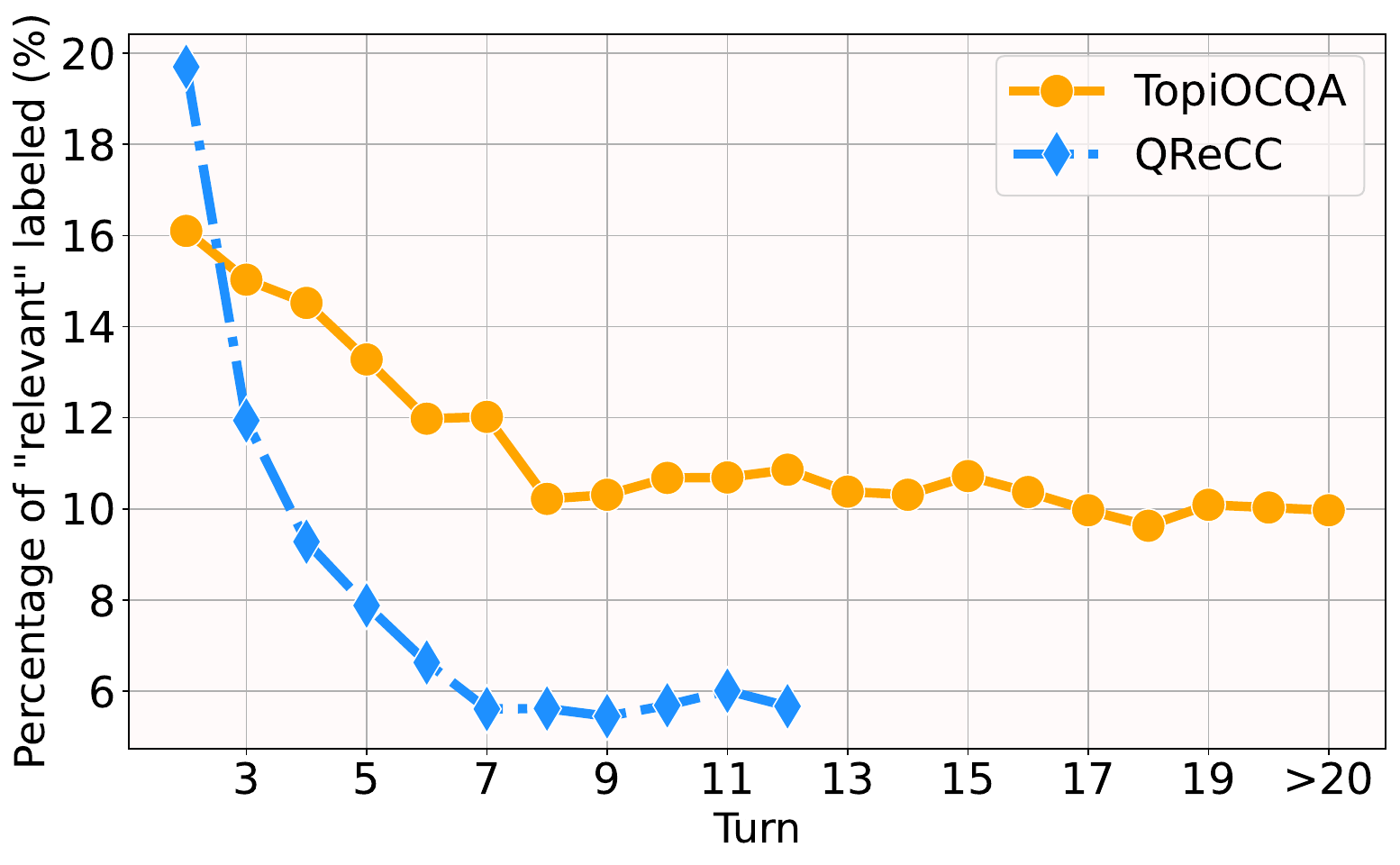}
    \caption{Portion of relevant historical turns over all historical turns, as conversations evolve.}
	\label{fig: PRJ}
\vspace{-2ex}
\end{figure}

The PRJs of historical turns are the foundation of context-denoised query reformulation and history-aware contrastive learning. 
The ablation study in Sec.~\ref{subsec: ablation} has shown the effectiveness of the approach. In this section, we take a deeper look to reveal the reasons behind the performance gain. Specifically, for a given turn ID $n$, we pool all historical turns in the dataset and compute the percentage of relevant ones as deemed by PRJs. Intuitively, this number denotes, on average, the portion of relevant historical turns over all historical turns. We plot this number against the turn ID in Figure~\ref{fig: PRJ}.

We observe that, overall, the relevant historical turns are only a fraction of all historical turns (up to 20\%). This verifies the necessity to perform PRJ for historical turns for context-denoised query reformulation. 
In addition, we see that the portion of the relevant history of TopiOCQA is generally greater than that of QReCC. This shows that our approach is reacting well to the abundant topic-switch phenomena in TopiOCQA. The PRJs derived from the topic-switches become the source of effectiveness for context-denoised query reformulation and history-aware contrastive learning, which finally results in pronounced gains on TopiOCQA.

Interestingly, the curves of both datasets show an intriguing trend of decrease-then-plateau. In the decreasing region, the amount of relevant history information does not scale as fast as the conversation. This shows the first several rounds of interactions have concentrated dependency on history. In contrast, as the conversation evolves, the amount of relevant history grows proportionally with the conversation (resulting in a plateaued percentage), which indicates a consistent and wide-spread dependency on history. We believe this insight on the change of history dependency over turns can inform future design of history modeling approaches. 

\begin{figure}[!t]
\centering
\includegraphics[width=0.8\linewidth]{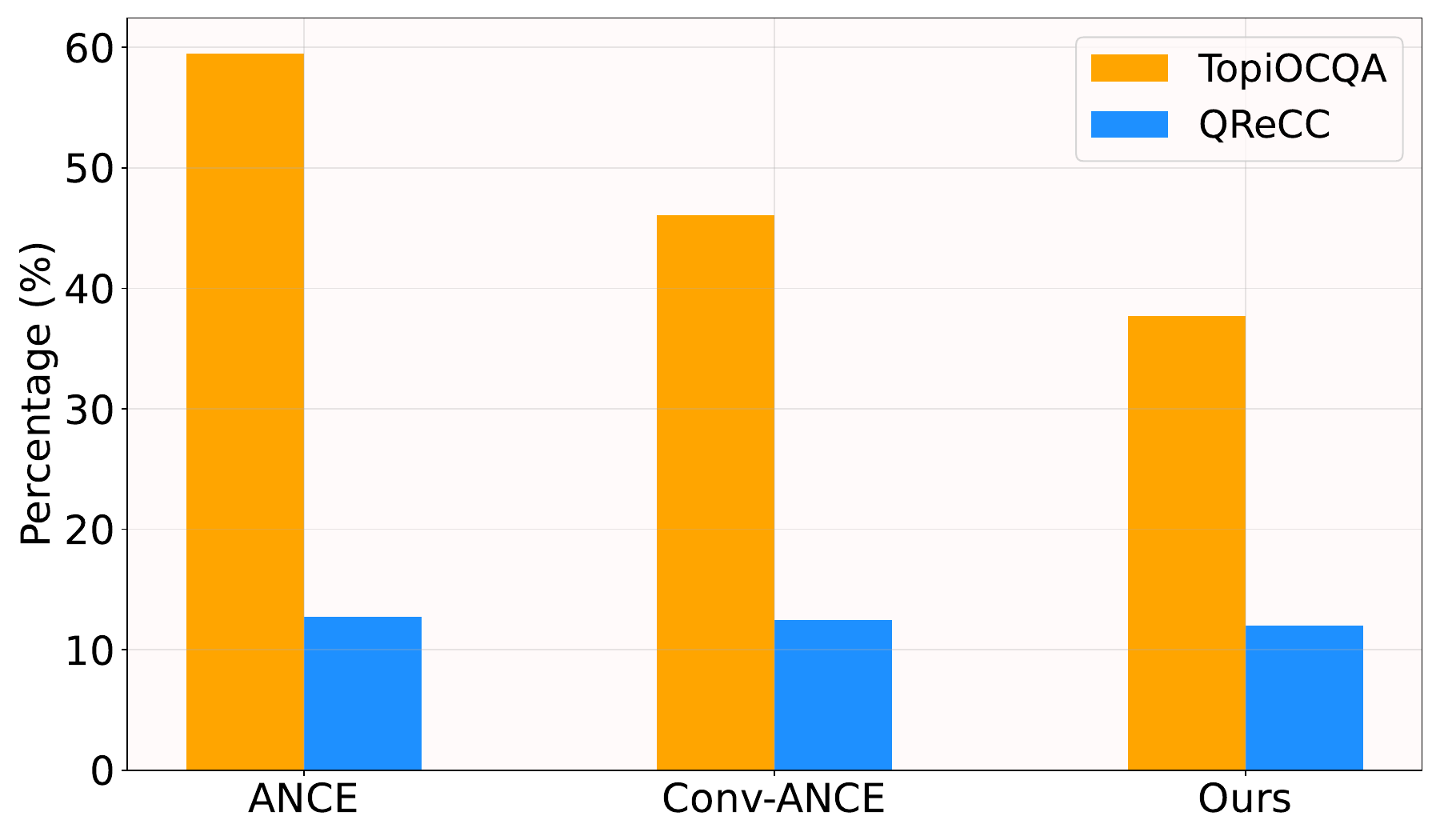}
	\caption{The percentage of the queries whose retrieved list has the ground-truth passage of the historical turns ranked higher than its own.}
	\label{fig: his}
\vspace{-2ex}
\end{figure}

\begin{table*}[!t]
\centering
\small
\begin{tabular}{lccccccccc}
\toprule
\multirow{2}{*}{Method} & \multirow{2}{*}{$k$} & \multicolumn{4}{c}{TopiOCQA} & \multicolumn{4}{c}{QReCC} \\
\cmidrule(lr){3-6}\cmidrule(lr){7-10}
~ & ~ & {MRR} & {NDCG@3} & {R@10} & {R@100} & {MRR} & {NDCG@3} & {R@10} & {R@100}\\
\midrule
\multirow{3}{*}{QR w/o PRJ} & 1 & 22.66 & 21.14 & 39.57 & 61.21 & 40.62 & 37.66 & 59.24 & 79.61\\
~ & 2 & 20.36 & 18.81 & 36.51 & 59.22 & 38.55 & 36.97 & 56.55 &  77.88\\
~ & 3 & 17.45 & 16.03 & 32.02 & 56.60 & 36.94 & 35.43 & 53.67 &  76.12\\
\midrule
\multirow{3}{*}{QR w/ PRJ} & 1 & 24.98 & 23.09 & 43.00 & \textbf{65.43} & 42.34 & 38.87 & 60.52 & 81.29\\
~ & 2 & 23.54 & 22.00 & 41.28 & 63.92 & 40.61 & 37.66 & 58.32 &  80.17\\
~ & 3 & 21.96 & 20.43 & 38.51 & 62.13 & 38.83 & 36.79 & 55.96 &  78.92\\
\midrule
Full model & 1 & \textbf{25.94} & \textbf{24.32} & \textbf{43.12} & 65.04 & \textbf{43.65} & \textbf{41.84} & \textbf{63.76} & \textbf{83.87}\\
\bottomrule
\end{tabular}
\caption{Performance on TopiOCQA and QReCC for the adapted use case of historical ground-truth passage substitution. The $k$ denotes the top-k passages in pseudo relevance feedback.}
\label{table: PRF}
\vspace{-2ex}
\end{table*}

\subsection{Impact of Historical Supervision Signals}
\label{subsec: impact of historical supervision signals}
We analyze how HAConvDR alleviates the issue of models favoring retrieving historical turn passages over current ones by examining the effect of historical supervision signals. \\

\noindent\textbf{Quantitative analysis} \quad The quantitative analysis is presented in Figure~\ref{fig: his}, which shows the percentage of the queries that rank the historical ground-truth passages higher than that of the current turn. 
We observe that our model can decrease the percentage of 
irrelevant historical gold passages for TopiOCQA, but not much for QReCC.
It indicates that the supervision signals for history-aware contrastive learning are stronger in TopiOCQA than in QReCC and it is consistent with the observation in Sec.~\ref{subsec: Main Results} that the improvements in TopiOCQA are more obvious. \\

\begin{figure}[t]
\centering
\includegraphics[width=1\linewidth]{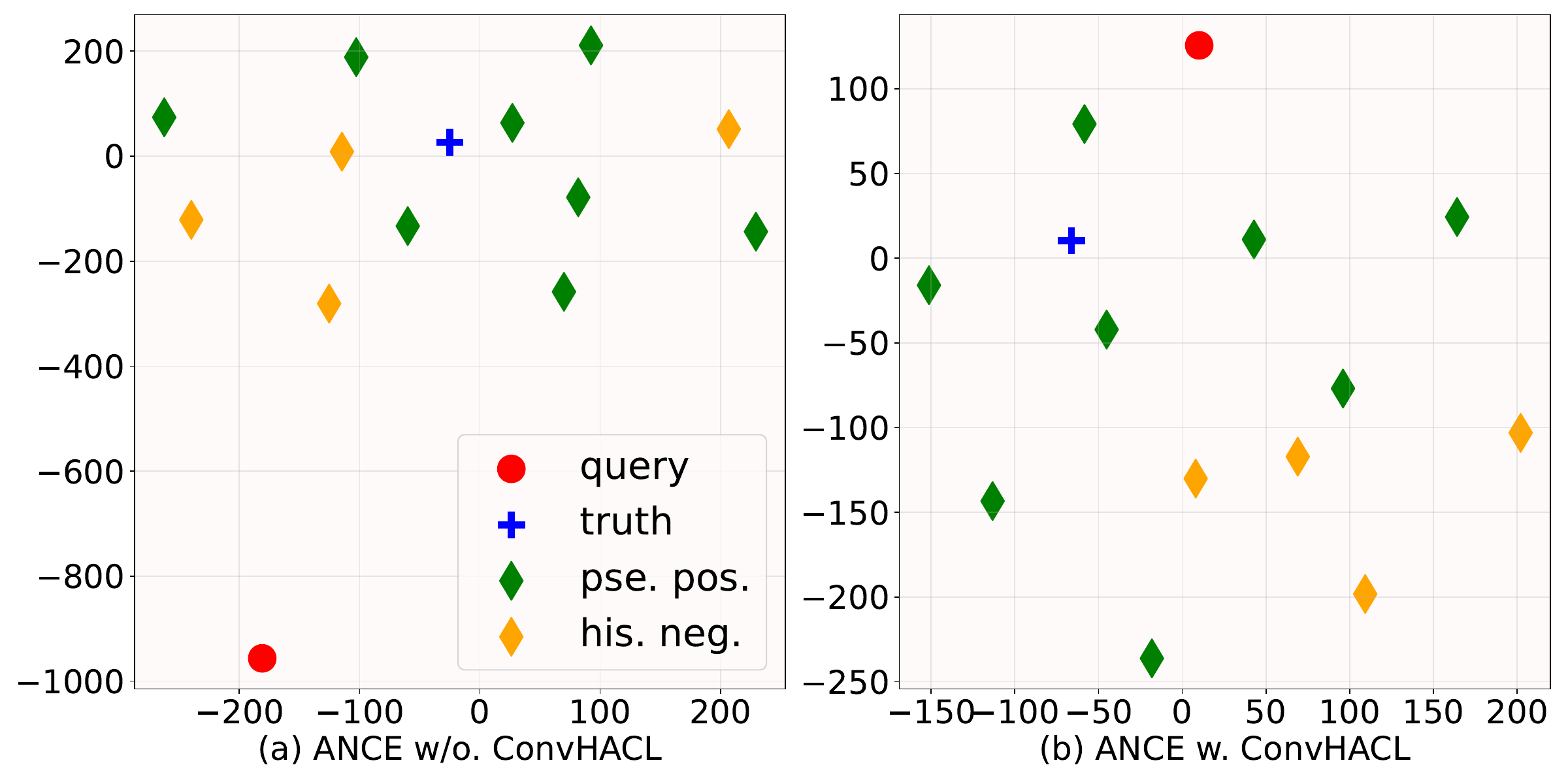}
	\caption{T-SNE visualization of query, ground-truth passage, and pseudo positives and history hard negatives embeddings via two ANCE models with and without HAConvDR training.} 
	\label{fig: tsne}
\end{figure}

\noindent \textbf{Qualitative analysis} \quad
To gain more insights into our approach, we did a qualitative study to visualize an example in the embedding space as 
Figure~\ref{fig: tsne}, which shows T-SNE visualization~\cite{van2008visualizing} to compare ANCE dense retriever with and without HAConvDR training in the embedding space. 
In contrast to the vanilla ANCE, which is unsuccessful in distinguishing the gold passage from the ground-truth of the historical turns, the ANCE trained with our HAConvDR exhibits a stronger ability to differentiate it from the distractors.
Besides, our model can also discriminate the gold passages of relevant and irrelevant turns, showing the effectiveness of these supervision signals toward better search results. The corresponding example is provided in Appendix~\ref{appendix: Qualitative Examples}.

\subsection{Impact of Substituting Historical Ground-Truth Passages}
\label{subsec: Applicability of PRF}
The computation of PRJs for historical turns relies on having access to historical ground-truth passages $\{p_i^*\}_{i=1}^{n-1}$. In many real-world applications, identifying ground-truth passages can be accomplished by analyzing user clicks, engagement, and feedback. However, we acknowledge that there are applications where historical ground-truth passages are difficult to obtain. In such cases, we can use top-retrieved passages as a substitute. This simple substitution allows us to perform the proposed approach described in Sec.~\ref{sec: method} with only minor modifications. Specifically, in Alg.~\ref{alg: PRJ} and Eq.~\ref{eq: reformulated query}, $p_i^*$ is approximated by the concatenation of the top-$k$ retrieved passages for $q_i$, where $k$ is a hyper-parameter. This retrieval is completed with the same backbone model of the conversational dense retriever. Meanwhile, $\mathcal{P}^{-}_{n}$ in Eq.~\ref{eq: pos neg sample} degrades to $\mathcal{P}^{-}_{b} \cup \mathcal{P}^{-}_{r}$. The rest of the approach is kept as is.

We conduct an ablation study to verify the effectiveness of our approach under this adaptation, with results presented in Table~\ref{table: PRF}.
We see the PRJ information still contributes to the retrieval performance of the reformulated query, further indicating its effectiveness. Besides, we find model performance degrades as $k$ increases, suggesting that longer contexts are more likely to contain noise, which cannot be entirely compensated by our approach. This suggests the potential for more advanced context-denoising approaches.
Finally, we find that using the full model with history-aware contrastive learning under the adapted setting continues to yield better results on top-ranking positions and outperforms most existing systems in Table~\ref{table: Main Results}.


\section{Conclusion}
In this paper, we present a new history-aware contrastive learning strategy for conversational dense retriever training, HAConvDR, which is based on context-denoised query reformulation and additional supervision signals mining from historical turns. 
Extensive experimental results on public datasets demonstrate the effectiveness of our model.
Furthermore, we conduct comprehensive analyses to gain insights into the impact of each component of HAConvDR on enhancing search performance and provide valuable insights on how they can work well for conversations with topic shifts.

\section*{Limitations}
Our work demonstrates the feasibility of using historical ground-truth passages for query reformulation and contrastive supervision signals.
Within our proposed HAConvDR, the context used for query reformulation includes selected historical passages, which are usually longer than hundreds of tokens. Thus, an explicit selection mechanism on raw text or an implicit fusion method on the latent representation could be designed to reduce the risk of information loss and the effect of noise.
Besides, an LLM-aided mechanism could be designed for query reformulation, e.g., selecting part of each historical passage that is helpful and with less noise as better supervision signals.
In addition, the historical supervised signals for model training might not be as important as the original annotation. Thus, a regulatory mechanism can be added to adjust the weight for pseudo positives within the history-aware conversational dense retrieval.

\section*{Acknowledgements}
This work is supported by a discovery grant from the Natural Science and Engineering Research Council of Canada and a Talent Fund of Beijing Jiaotong University (2024JBRC005).

\bibliography{anthology,custom}
\bibliographystyle{acl_natbib}

\appendix

\section{More Detailed Experimental Setup}

\subsection{Datasets}
\label{appendix: datasets}
The statistics of each dataset are presented in Table~\ref{table:datasets} where we eliminate the samples without gold passages in QReCC.
The details of each dataset are in the following: \\

\noindent \textbf{TopiOCQA} addresses the novel issue of topic switching, a common occurrence in realistic scenarios. In typical conversations, there are usually over 10 turns and a minimum of 3 topics. Furthermore, turns related to the same topic tend to have similar gold passages, thus we could leverage them as additional supervision signals. \\

\noindent \textbf{QReCC} primarily addresses the task of query rewriting by attempting to reformulate the query to approach the human-rewritten query. In comparison to TopiOCQA, QReCC involves conversations with a smaller number of turns, and most of these conversations revolve around the same topic. As a result, turns within the same conversation often yield identical gold passage results, making it possible to extract only a limited number of additional supervision signals.

\begin{table}[!ht]
\centering
\small
\setlength{\tabcolsep}{4pt}{
\begin{tabular}{llrrr}
\toprule
Dataset & Split & \#Conv. & \#Turns(Qry.) & \#Collection \\ \midrule
\multirow{2}{*}{TopiOCQA} & Train & 3,509 & 45,450 & \multirow{2}{*}{25M} \\
 & Test  & 205 & 2,514 & \\
\midrule
\multirow{2}{*}{QReCC} & Train & 10,823 & 29,596 & \multirow{2}{*}{54M} \\
 & Test  & 2,775 & 8,124 & \\
\bottomrule
\end{tabular}}
\caption{Statistics of conversational search datasets.}
\vspace{-2ex}
\label{table:datasets}
\end{table}

\subsection{Implementation Details}
\label{appendix: Implementation}
We implement all models by PyTorch~\cite{DBLP:conf/nips/PaszkeGMLBCKLGA19} and Huggingface's Transformers~\cite{DBLP:journals/corr/abs-1910-03771}. 
The experiments are conducted on one Nvidia A100 40G GPU. For conversational dense retriever training, we use Adam optimizer with 3e-5 learning rate and set the batch size as 32. The maximum length of the reformulated query and the passage as model input is 512 and 384 for TopiOCQA and both 256 for QReCC, respectively. For the compared baseline systems, we implement the main competitor SDRConv with the same number of hard negatives and batch size as ours and use the ANCE+InstructoR$_\text{QRPG}$ version in InstructoR for fair comparison. All the dense retrievers are initiated with ANCE.
For evaluation, We adopt the \texttt{pytrec\_eval} tool~\citep{sigir18_pytrec_eval} for metric computation.\\

\section{Qualitative Example}
\label{appendix: Qualitative Examples}
Table~\ref{table: Qualitative Example} presents a qualitative example corresponding to the T-SNE visualization in Figure~\ref{fig: tsne}, which gives a comprehensive understanding of how historical ground-truth passage can benefit current query retrieval as supervision signals.
\begin{table*}[t]
    \centering
    \begin{tabular}{p{15cm}}
    \toprule
    \textbf{Conversation (id:4-13)}\\
    \midrule
    $q_1$: who sang all i want for christmas in 1995? (irrelevant)\\
    $p_1$: All I Want for Christmas Is You is a Christmas song by American singer-songwriter ... (536, -, -)\\
    $q_2$: who is she? (relevant)\\
    $p_2$: Mariah Carey (born March 27, 1969 or 1970) is an American singer-songwriter ... (5, 20, 17)\\
    $q_3$: what was her early days like? (irrelevant)\\
    $p_3$: Mariah Carey was born in Huntington, New York, on March 27, 1969 or 1970 ... (614, -, -)\\
    $q_4$: what are some famous songs she performed during 2010? (relevant)\\
    $p_4$: It missed out on the top one-hundred in the United Kingdom by one position ... (-, -, -)\\
    $q_5$: who composed the former mentioned one? (irrelevant)\\
    $p_5$: Cox plated the keyboard and percussion. The background vocals were sung by ... (-, -, -)\\
    $q_6$: how did it perform in the charts? (relevant)\\
    $p_6$: In the United States, Oh Santa! became a record-breaking entry on ... (-, -, -)\\
    $q_7$: how was it received critically? (relevant)\\
    $p_7$: Mike Diver of the BBC wrote that Oh Santa! is a ``boisterous'' song ... (-, -, -)\\
    $q_8$: what was her other song about? (irrelevant)\\
    $p_8$: Auld Lang Syne (The New Year's Anthem) is a re-write of Auld Lang Syne ... (-, -, -)\\
    $q_9$: how was it received critically? (relevant)\\
    $p_9$: Auld Lang Syne (The New Year's Anthem) garnered a negative response from critics ... (937, -, 322)\\
    $q_{10}$: what are some philanthropic activities this singer is associated with? (relevant)\\
    $p_{10}$: Carey is a philanthropist who has been involved with several ... (197, 502, 31)\\
    $q_{11}$: what does the latter mentioned foundation do? (relevant)\\
    $p_{11}$: The Make-A-Wish Foundation is a 501(c)(3) nonprofit organization founded in ... (-, -, -)\\
    $q_{12}$: what is her style of music? (relevant)\\
    $p_{12}$: Love is the subject of the majority of Carey's lyrics, although she has written ... (6, 68, 29)\\
    $q_{13}$: what are some awards she has received?\\
    \midrule
    \textbf{Gold Passage} (107, 68, 2) \\
    \midrule
    Throughout her career, Carey has earned numerous awards and honors, including the World Music Awards', Best Selling Female Artist of the Millennium, the Grammy Award for Best New Artist in 1991, and \"Billboard\"s Special Achievement Award for the Artist of the Decade during the 1990s. In a career spanning over 20 years,  Carey has sold over 200 million records worldwide, making her one of the best-selling music artists of all  time. Carey is ranked as the best-selling female artist of the Nielsen SoundScan era, with over 52 million copies sold.  Carey was ranked first in MTV and \"Blender\" magazine's 2003 countdown of the 22 Greatest Voices in Music, and was placed second in \"Cove\" magazine's list of \"The 100 Outstanding Pop Vocalists.\" Aside from her voice, she has become known for her songwriting.\\
    \bottomrule
    \end{tabular}
     \caption{A qualitative example of how historical ground-truth passage can benefit current query retrieval as supervision signals within HAConvDR. The brackets following each historical query indicate whether it is relevant or irrelevant to the current turn. The brackets with three numbers after each historical gold passage indicate its rank position by ANCE, Conv-ANCE, and our HAConvDR within top-1000, where ``-'' means it is ranked outside the top-1000.}
     \label{table: Qualitative Example}
\end{table*}

\end{document}